\documentclass[12pt]{article}
\usepackage{graphicx,amssymb, amsmath}
\newcommand{\mathsym}[1]{{}}

\input epsf.tex

\let\badcite=\cite
\def\cite{~\badcite}

\def\slashchar#1{\setbox0=\hbox{$#1$}           
   \dimen0=\wd0                                 
   \setbox1=\hbox{/} \dimen1=\wd1               
   \ifdim\dimen0>\dimen1                        
      \rlap{\hbox to \dimen0{\hfil/\hfil}}      
      #1                                        
   \else                                        
      \rlap{\hbox to \dimen1{\hfil$#1$\hfil}}   
      /                                         
   \fi}
    \def\slashword#1{\setbox0=\hbox{$#1$}        
  \dimen0=\wd0                                   
   \setbox1=\hbox{/} \dimen1=\wd1                
   \ifdim\dimen0>\dimen1                         
      \rlap{\hbox to \dimen0{\hfil\bf---\hfil}} %
      #1                                         %
   \else                                         
      \rlap{\hbox to \dimen1{\hfil$#1$\hfil}}    
      /                                          
    \fi}                                         %

\catcode`@=11
\newdimen\vbigd@men                             

\def\vbig#1#2{{\vbigd@men=#2\divide\vbigd@men by 2%
   \hbox{$\left#1\vbox to \vbigd@men{}\right.\n@space$}}}
\catcode`@=12

\catcode`@=11
\def\citenum#1{\csname b@#1\endcsname}
\catcode`@=12

\begin{document}
\begin{titlepage}

\begin{flushright}
{SCUPHY-TH-08004}\\
{CAS-KITPC/ITP-083}\\
\end{flushright}

\bigskip\bigskip

\begin{center}{\Large\bf\boldmath
On Measuring Split-SUSY Neutralino and Chargino Masses at the LHC }
\end{center}
\bigskip
\centerline{\bf N. Kersting\footnote{Email: nkersting@scu.edu.cn} }
\centerline{{\it Physics Department, Sichuan University, P.R. China 610065}}
\centerline{{\it and}}
\centerline{{\it Kavli Institute for Theoretical Physics, Beijing, P.R. China 100086}}

\bigskip

\begin{abstract}
In Split-Supersymmetry models, where the only non-Standard Model states produceable at LHC-energies consist of a  gluino plus
neutralinos and charginos, it is conventionally accepted that only mass \emph{differences} among these latter are measureable at the LHC. The present work shows that application of a simple `Kinematic Selection' technique allows full reconstruction of neutralino and chargino masses from one event, in principle. A Monte Carlo simulation demonstrates the feasibilty of using this technique at the LHC.
\end{abstract}

\newpage
\pagestyle{empty}

\end{titlepage}


\section{Introduction}

As data from the LHC is recorded and analyzed in the upcoming years, experimentalists will look
for  signatures of physics beyond the Standard Model (SM), in particular Supersymmetry (SUSY). The most well-studied scenario,
in which SUSY  alleviates the hierarchy problem with a low-energy (sub-TeV) spectrum of sparticles, entails
copious production of strongly-interacting squarks and gluinos, identified via their associated jets\cite{atlas,cms}, which cascade through numerous decay channels involving other sparticles, including e.g.  electroweak (EW)-interacting sleptons, neutralinos, and charginos --- the masses of these sparticles, whose precise values are crucial to understanding features of an underlying fundamental theory,
 may be reconstructed (at least partially)  from measurements of various invariant mass endpoints in certain exclusive decay channels (e.g.\cite{invmass}). Of course it is also entirely possible that the SUSY spectrum is far above the TeV level, hence inaccessible at the LHC.
In between the above two extremes, phenomenologically speaking, is the scenario where some of the sparticles are light, while others extremely massive and decoupled,  ``Split SUSY"\cite{split-susy}  providing the most popular example.

At low energies, the Split-SUSY spectrum, aside from the established SM particles and one light Higgs boson, contains only four neutralinos ($\widetilde{\chi_i}^0$, $i=1..4$), two charginos ($\widetilde{\chi_j}^\pm$, $j=1,2$), and a long-lived gluino ($\tilde{g}$). The phenomenology of this latter, which would be expected to form so-called ``R-hadrons" in the detector (and therefore not immediately decaying to other sparticles) has been thoroughly covered elsewhere (see\cite{gluinos, Kilian}) and will not concern us here\footnote{Even if the gluino does decay in some corner of parameter space, this will only assist with the current study by boosting signal rates.}. The focus of the present study is rather on the neutralinos and charginos, hereafter collectively referred to as `EW-inos'.
These cannot decay via squarks and sleptons, which are many orders of magnitude heavier, but  must rather decay promptly via a Higgs or EW gauge bosons ($Z^0$, $W^\pm$). If mass differences between EW-inos are smaller than $m_Z$ or $m_W$, they will undergo 3-body decays to quarks or leptons: e.g. ${\widetilde\chi_{i}}^0 \to {Z^0}^*(\to \ell^\pm \ell^\mp,~ q \overline{q}) {\widetilde\chi_1}^0$, ${\widetilde\chi_{2}}^\pm \to {Z^0}^*(\to \ell^\pm \ell^\mp,~ q \overline{q}) {\widetilde\chi_1}^\pm$, and ${\widetilde\chi_1}^\pm \to {W^\pm}^*(\to \ell^\pm \nu,~ q q') {\widetilde\chi_1}^0$.
In particular, dilepton pairs from the above off-shell Z decays will have an invariant mass distribution which cuts off sharply at
$m_{\widetilde{\chi}_{i}^0} - m_{\widetilde{\chi}_{1}^0}$ or $m_{\widetilde{\chi}_{2}^\pm} - m_{\widetilde{\chi}_{1}^\pm}$, so the usual conclusion from Split-SUSY studies is that only mass \emph{differences} between EW-inos are measurable at the LHC\cite{Kilian}.

The thesis of this work is that one can do much better than just find EW-ino mass differences --- the masses themselves can be reconstructed from additional kinematic analysis made possible by the fact that EW-inos must be pair-produced to preserve R-parity, which is theoretically motivated in making the lightest SUSY particle (LSP)  a good dark matter candidate. Each SUSY EW-ino event
($\widetilde{\chi_{i}}^\pm \widetilde{\chi_{j}}^\mp$, $\widetilde{\chi_{i}}^\pm \widetilde{\chi_{j}}^0$, or $\widetilde{\chi_{i}}^0 \widetilde{\chi_{j}}^0$) may thus contain multiple (as many as ten) hard leptons, the momenta of which, when contracted into all possible invariant masses (as in\cite{Huang1}), encode much information.
This was previously overlooked in the literature, presumably because pairs of EW-inos arising from hadronic collisions carry an uncertain center-of-mass energy, hence yielding final state leptons not amenable to the usual invariant-mass endpoint analysis.
As shown in Hidden Threshold (HT) methods\cite{Huang2}, however,  correlations among such invariants (i.e. a Dalitz plot) still
carry information about endpoints and, more importantly for the present work, distribute events according to the kinematics in each respective EW-ino decay frame. The strategy in this work, therefore, is to focus on one region of a Dalitz plot where events must arise from the same decay frame kinematics, find the Lorentz-boosts back to the frames of the decaying EW-inos, and match energies/momenta (including measured missing transverse momenta) to extract the values of relevant masses. A single (perfect) event may suffice for full reconstruction in principle, though in practice (including  detector effects and backgrounds) one must do with a collection of less-than-perfect events which will give statistical distributions of the unknown masses.

In the following, let us then proceed thusly: Section \ref{sec:app} will explain this `Kinematic Selection' method in the context of Split-SUSY EW-ino decays; Section \ref{sec:lorentz} will detail how to reconstruct EW-ino masses from a perfect event and Section \ref{sec:mc} will then test and confirm the feasibility of this in a Monte Carlo simulation appropriate to the LHC environment. Summary and comments on further elaborations are contained in Section \ref{sec:disc}.

  \section{Kinematic Selection Technique for EW-inos}
  \label{sec:app}
  In Split-SUSY models, EW-inos at the LHC can only be pair-produced in quark-quark s-channel processes through an off-shell $W$, $Z$, or $\gamma$:
  \begin{eqnarray}\label{prod1}
     qq' \to W^* &\to& {\widetilde\chi_i}^\pm {\widetilde\chi_j}^0 \\ \label{prod2}
     q \overline{q} \to Z^*  &\to& {\widetilde\chi_i}^0 {\widetilde\chi_j}^0 \\ \label{prod3}
       q \overline{q} \to Z^* / \gamma^* &\to& {\widetilde\chi_i}^\pm {\widetilde\chi_j}^\mp
\end{eqnarray}
Then, since each EW-ino cannot decay through squarks or sleptons, but only through a $Z^0$ or $W^\pm$ (or a light higgs $h^0$, though this tends to be subdominant),  it must decay among the following five tree-level\footnote{Loop-level decays can also have phenomenological importance, e.g. $\widetilde{\chi_{i}}^0 \to \gamma \widetilde{\chi_{1}}^0$\cite{Cheung}.} channels (taking  $m_{\widetilde{\chi_1}^\pm} > m_{\widetilde{\chi_2}^0}$ and $m_{\widetilde{\chi_2}^\pm} > m_{\widetilde{\chi_4}^0}$):
  \begin{eqnarray} \label{gdecay1}
    \widetilde{\chi_{2}}^\pm &\to& Z^0 \widetilde{\chi_{1}}^\pm\\
   \widetilde{\chi_{2}}^\pm &\to& W^\pm \widetilde{\chi_{i}}^0 ~~~~~~~ (i=1..4) \\
   \widetilde{\chi_{1}}^\pm &\to& W^\pm \widetilde{\chi_{i}}^0   ~~~~~~~ (i=1,2)\\
    \widetilde{\chi_{i}}^0 &\to& Z^0 \widetilde{\chi_{j}}^0  ~~~~~~~ (i=2..4),~~ (j=1..i-1) \\
    \widetilde{\chi_{i}}^0 &\to& W^\pm \widetilde{\chi_{1}}^\mp ~~~~~~~ (i=3,4) \label{gdecay5}
  \end{eqnarray}
where the $Z^0$ or $W^\pm$ could be on- or off-shell. The number of possible decay chains combining (\ref{prod1})-(\ref{prod3}) and (\ref{gdecay1})-(\ref{gdecay5}), even without distinguishing on- or off-shell intermediaries or considering the rest of the decay chain,  is already quite large, but most of these, fortunately,  will not be needed in the present study.

  \subsection{Chargino-Neutralino Modes}

The most heavily-produced state in Split SUSY models is likely to be a
chargino-neutralino pair ${\widetilde\chi_1}^\pm {\widetilde\chi_2}^0$, since these sparticles are relatively light and well-mixed,
where it is further assumed that they proceed to decay through an offshell\footnote{If decays occur through on-shell $Z^0$ and $W^\pm$, the
signal is much more challenging to extract, being swamped by $WZ$ and $ZZ$ backgrounds. } $Z^0$ or $W^\pm$ to leptons ($\ell = e, \mu$):
\begin{eqnarray} \label{w1decay}
  {\widetilde\chi_1}^\pm &\to& {W^\pm}^*(\to \ell^\pm \nu) {\widetilde\chi_1}^0  \\
  {\widetilde\chi_{2}}^0 &\to& {Z^0}^*(\to \ell^\pm \ell^\mp) {\widetilde\chi_1}^0 \label{z2decay}
\end{eqnarray}
The endstate will therefore contain three leptons (of which at least two are opposite-sign-same-flavor (OSSF)) whose momenta $p_{1,2,3}$ can, in the spirit of\cite{Huang1}, be systematically contracted into three independent invariant masses\footnote{These have the advantage of systematic definition and symmetry under lepton interchange at the cost of algebraic complexity. }:
 \begin{eqnarray}\label{av3l}
   \overline{M}_{{3l}}^2 &\equiv&
  (p_1 + p_{2}+ p_3)^2  \\
  \overline{M}_{{l2l}}^4 &\equiv&  \{
  (p_1 + p_{2}- p_3)^4 +
  (p_3 + p_{1}- p_2)^4 +
  (p_{2}+ p_3-p_1)^4  \}/3
    \\  \label{avl2l}
  \overline{M}_{{ll}}^4 &\equiv&
   \{
  (p_1 + p_{2})^4 +
   (p_1 + p_{3})^4
  + ( p_2 + p_{3})^4  \} /3
  \label{avll}
\end{eqnarray}

  The problem now, in which the HT technique assists, is how to use the information contained in the above invariants to select events with a desired kinematic configuration.
   Observe, first of all, that the off-shell W in (\ref{prod1}) will itself have an invariant mass somewhere (and unpredictably) in the range
  $m_{\widetilde{\chi}_{1}^\pm} + m_{\widetilde{\chi}_{2}^0} < m^* < E_{LHC}$ where $E_{LHC} \sim 14\, \hbox{TeV}$ is the theoretical maximum partonic collision energy at the LHC. Let us consider the case where $m^* = m_{\widetilde{\chi}_{1}^\pm} + m_{\widetilde{\chi}_{2}^0}$ and designate this `threshold production'.

\begin{figure}[!htb]
\begin{center}
\includegraphics[width=2.5in]{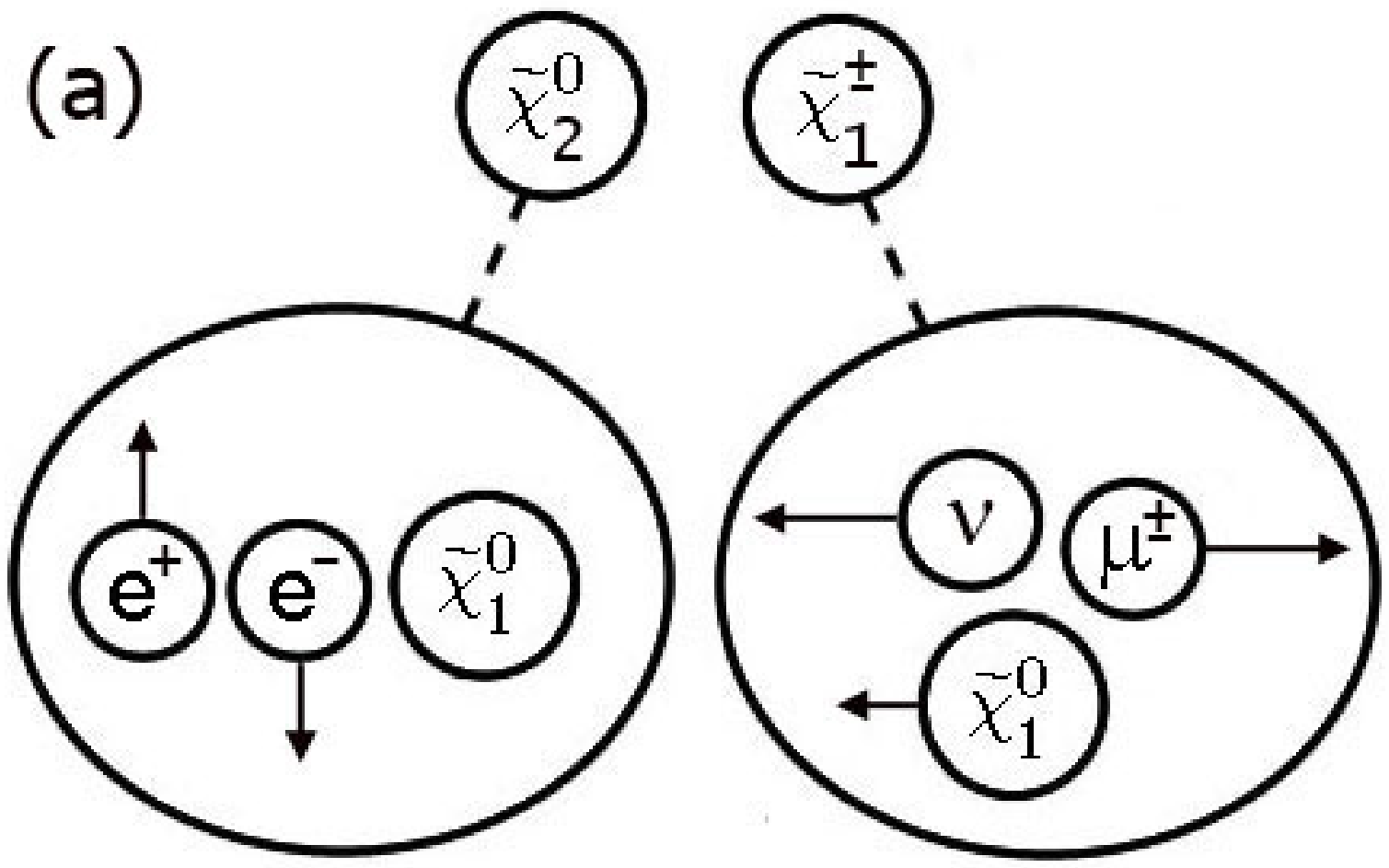}
\includegraphics[width=2.5in]{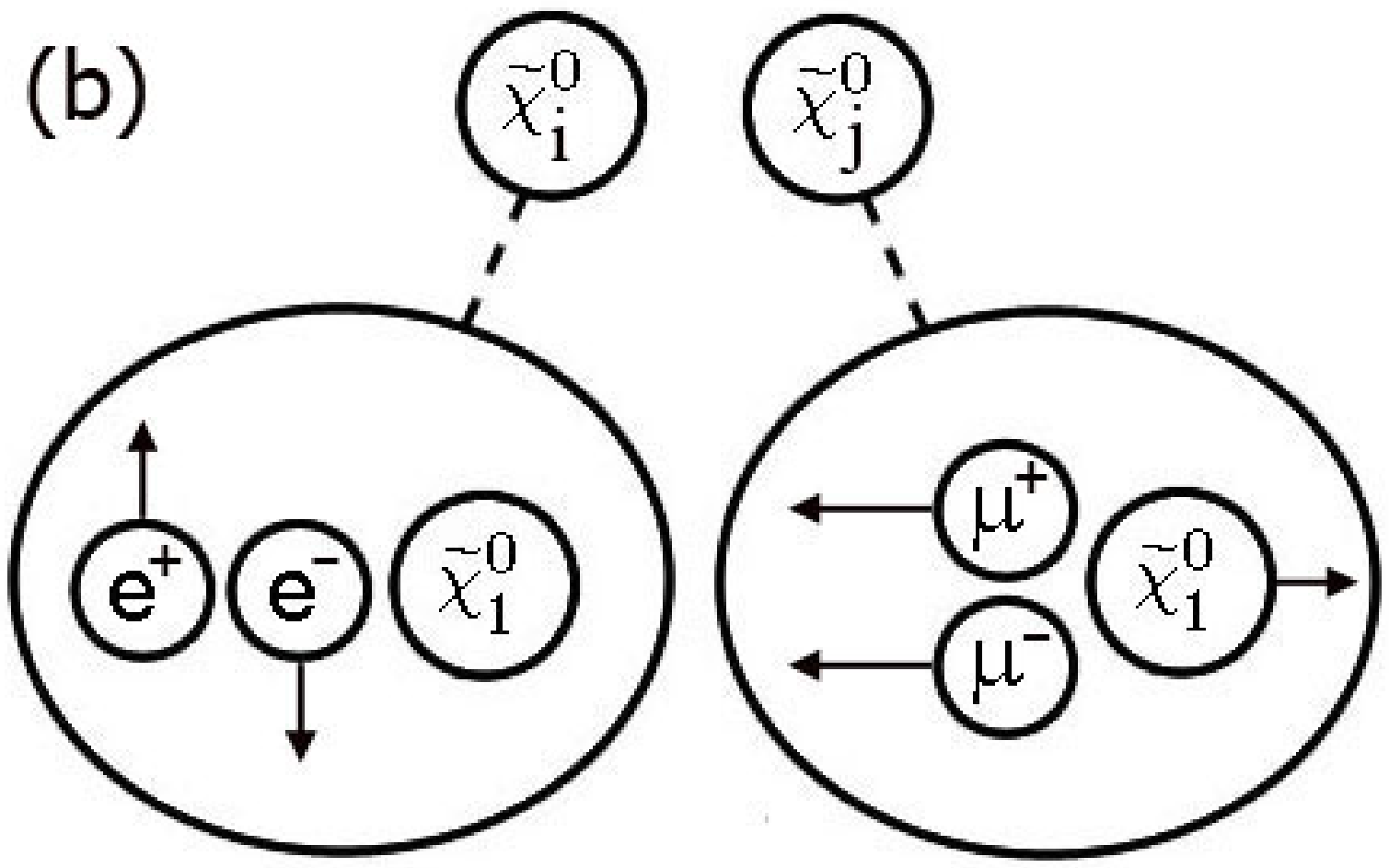}
\end{center}
\caption{\small \emph{Schematic of the kinematic configurations (with $e \leftrightarrow \mu $ as well) which minimize various invariant masses at threshold  for
(a) Chargino-Neutralino and  (b) Neutralino-Neutralino modes (these particles decay at rest in this frame). }
}
 \label{fig:3body}
\end{figure}

To simplify the discussion, assume we have a $e^+ e^- \mu^\pm$ endstate (the following will also pertain to same flavor states $e^+ e^- e^\pm$ or $\mu^+ \mu^- \mu^\pm$ with the correct lepton-pairing). From relativistic kinematics, it is quite straigtforward to show that, for threshold production, when
$M_{e^+ e^-} \equiv (p_{e^+} + p_{e^-})^2$ is \emph{maximal}, $\overline{M}_{{l2l}}$ is \emph{minimal} when the kinematical configuration in Fig.~\ref{fig:3body}a is attained: in the rest frame of the parents $\widetilde{\chi}_{1}^\pm$ and
$\widetilde{\chi}_{2}^0$, the electron and positron are produced back-to-back with maximal momentum along  directions perpendicular to the muon, which also carries maximal momentum (see Appendix for derivation). The minimal value of $\overline{M}_{{l2l}}$
  is then given by
\begin{equation}\label{ml2lmin}
    \overline{M}_{l2l}^{min} = \sqrt{m_{\widetilde{\chi}_{2}^0} - m_{\widetilde{\chi}_{1}^0}}
    \left( \frac{
    2(m_{\widetilde{\chi}_{2}^0} - m_{\widetilde{\chi}_{1}^0})^2
    + (m_{\widetilde{\chi}_{1}^\pm} - m_{\widetilde{\chi}_{2}^0} + m_{\widetilde{\chi}_{1}^0}
   -  m_{\widetilde{\chi}_{1}^0}^2 / m_{\widetilde{\chi}_{1}^\pm})^2  }{3}
       \right) ^{1/4}
\end{equation}
(the other invariants ${M}_{{3l}}$ and $\overline{M}_{{ll}}$ are, on the contrary, trivially minimized by $p_{\mu^\pm} = 0$,
giving  ${M}_{{3l}}^{min} = M_{e^+ e^-}^{max}$  and $\overline{M}_{ll}^{min} = (1/3)^{1/4}M_{e^+ e^-}^{max}$ , respectively).
For the more realistic case of non-threshold production, i.e. $m^* > m_{\widetilde{\chi}_{1}^\pm} + m_{\widetilde{\chi}_{2}^0}$ giving a relative velocity $\overrightarrow{\beta}$ between the $\widetilde{\chi}_{1}^\pm$ and $\widetilde{\chi}_{2}^0$, it is also easy to show that $M_{e^+ e^-}$ is maximal and $\overline{M}_{{l2l}}$ minimal (for \textit{this} $\beta$) for the same  decay configuration (in the respective $\widetilde{\chi}_{1}^\pm$ and $\widetilde{\chi}_{2}^0$ `parent-frames')  of Fig.~\ref{fig:3body}a.
 The relevant Dalitz plot is therefore ``$M_{e^+ e^-}$ vs.  $\overline{M}_{{l2l}}$", the events of interest accumulating along where
the line $M_{e^+ e^-} = M_{e^+ e^-}^{max}$ ($= m_{\widetilde{\chi}_{2}^0} - m_{\widetilde{\chi}_{1}^0}$) hits the kinematically-allowed portion of the plot\footnote{One might hope to measure the endpoint (\ref{ml2lmin}) from the intersection and thereby constrain SUSY masses, but this turns out to require very high (sub-GeV) endpoint precision and is much inferior to the present method.}.

\subsection{Neutralino-Neutralino Modes}

Neutralino pair production (\ref{prod2})  may also be significant,
but only for \emph{unlike} neutralinos, i.e. $i \ne j$, due to a suppression
of the $Z^0{\widetilde\chi_i}^0{\widetilde\chi_i}^0$ coupling\cite{Kilian, Bian}.
Assuming again that the neutralinos decay through a 3-body decay as in (\ref{z2decay}), we
now have an endstate described by four lepton momenta $p_{1,2,3,4}$ which can be analyzed via a set of seven invariant masses\cite{Huang1}, including
e.g.
\begin{eqnarray}\label{4linv}
  M_{4l}^2 &\equiv& (p_1 + p_{2}+ p_3 + p_{4})^2 \\ \nonumber
  \overline{M}_{{2l2l}}^4 &\equiv& \{(p_1 + p_{2}- p_3 - p_{4})^4 +
  (p_1 + p_{4}- p_2 - p_{3})^4
  + (p_2 + p_{4}- p_{1} - p_{3})^4\} /3
\end{eqnarray}
Going through the same argument above for ${\widetilde\chi_1}^\pm {\widetilde\chi_2}^0$ modes, one finds
that the threshold kinematic configuration in Fig.~\ref{fig:3body}b, where $M_{e^+ e^-}$ is maximal, forces
$M_{4l}$ and $\overline{M}_{{2l2l}}$ to attain the minima (derived in Appendix)
\begin{eqnarray}\label{4lmins}
 {M}_{4l}^{min} &=& \sqrt{(m_{\widetilde{\chi}_{i}^0} - m_{\widetilde{\chi}_{1}^0})
 (m_{\widetilde{\chi}_{j}^0} + m_{\widetilde{\chi}_{i}^0} - m_{\widetilde{\chi}_{1}^0}
   -  m_{\widetilde{\chi}_{1}^0}^2 / m_{\widetilde{\chi}_{j}^0})
 } \\
  \overline{M}_{2l2l}^{min} &=& \sqrt{m_{\widetilde{\chi}_{i}^0} - m_{\widetilde{\chi}_{1}^0}}
    \left( \frac{
    2(m_{\widetilde{\chi}_{i}^0} - m_{\widetilde{\chi}_{1}^0})^2
    + (m_{\widetilde{\chi}_{j}^0} - m_{\widetilde{\chi}_{i}^0} + m_{\widetilde{\chi}_{1}^0}
   -  m_{\widetilde{\chi}_{1}^0}^2 / m_{\widetilde{\chi}_{j}^0})^2  }{3}
       \right) ^{1/4} \nonumber
       \end{eqnarray}
and events of this type can be found near these minima on a plot of  $M_{e^+ e^-}$ (or  $M_{\mu^+ \mu^-}$) versus $M_{4l}$ and $\overline{M}_{{2l2l}}$.

  \begin{figure}[!htb]
\begin{center}
\includegraphics[width=2.5in]{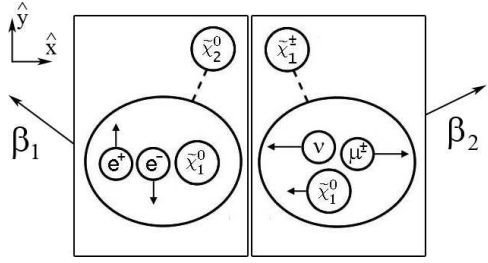}
\end{center}
\caption{\small \emph{ Class of parent-frame kinematics sought: in the frame where the $\widetilde{\chi}_{2}^0$ decays at rest, the $e^\pm$ are along global coordinates $\pm \hat{y}$ with maximal energy; likewise in the $\widetilde{\chi}_{1}^\pm$ decay frame the muon is along $\pm \hat{x}$ with maximal energy. Any velocities $\beta_{1,2}$ of these decaying states  in the lab frame are permitted.}
}
 \label{fig:3bodybeta}
\end{figure}

\section{Finding Masses from One Event}
\label{sec:lorentz}
Let us now suppose we have a $e^+ e^-\mu^\pm$ event of maximal $M_{e^+ e^-}$ and minimal $\overline{M}_{l2l}$ (a very similar analysis can be done for  $e^+ e^-\mu^+ \mu^-$ events from $\widetilde{\chi}_{i}^0 \widetilde{\chi}_{j}^0$ decays, but we will not need this for the present study\footnote{If the number of  $e^+ e^-\mu^+ \mu^-$ events is sufficiently high we can also get events
where $M_{e^+ e^-}$ and $M_{\mu^+ \mu^-}$ are both maximal, also allowing for a straightforward mass-reconstruction\cite{Kersting}. }). By the discussion of the last section the kinematic configuration must be of the type shown in  Fig.~\ref{fig:3bodybeta}, i.e. if the decaying $\widetilde{\chi}_{2}^0$ and $\widetilde{\chi}_{1}^\pm$ had no  motion
 then the $e^\pm$ would have equal and opposite momenta (along say $\pm \hat{y}$) while the muon would be emitted perpendicular to $\hat{y}$, say along $\hat{x}$, also with maximal kinetic energy; the $\widetilde{\chi}_{2}^0$ and $\widetilde{\chi}_{1}^\pm$ are, however, permitted to be moving with different velocities $\overrightarrow{\beta}_{1,2}$, so the observed leptonic momenta will generally point in random directions.

These leptonic momenta nevertheless carry useful information, for
if we knew $\overrightarrow{\beta}_{1,2}$ as well  we could reconstruct all three unknown masses $m_{\widetilde{\chi}_{1}^0}$,  $m_{\widetilde{\chi}_{2}^0}$, and  $m_{\widetilde{\chi}_{1}^\pm}$ as follows: the observed missing transverse momentum $\overrightarrow{\slashchar{p}_T}$  must clearly arise from the invisible particles $2\widetilde{\chi}_{1}^0 + \nu$, and since their 4-momenta are known functions of the masses in the $\widetilde{\chi}_{2}^0$ and $\widetilde{\chi}_{1}^\pm$ decay frames, it is a simple matter of Lorentz-boosting these by $\overrightarrow{\beta}_{1,2}$ to the lab frame and matching to the two components of $\overrightarrow{\slashchar{p}_T}$. Two matching conditions plus the dilepton edge
 ($M_{e^+ e^-}^{max} = m_{\widetilde{\chi}_{2}^0} - m_{\widetilde{\chi}_{1}^0}$) determines the set $\{m_{\widetilde{\chi}_{1}^0},~ m_{\widetilde{\chi}_{2}^0},~m_{\widetilde{\chi}_{1}^\pm}\}$.

It is easy, in fact, to determine $\overrightarrow{\beta}_1$: this corresponds to the unique Lorentz transformation
 $\mathbf{\Lambda}_1$ which makes the transformed
$e^\pm$ momenta equal and opposite ($\equiv \pm\overrightarrow{p}'$), as well as simultaneously bringing the corresponding $\widetilde{\chi}_{1}^0$ to rest, and is given by
\begin{equation}\label{betaeqn}
    \overrightarrow{\beta}_1 = \frac{\overrightarrow{p}_{e^+} + \overrightarrow{p}_{e^-}}{E_{e^+} + E_{e^-}}
\end{equation}
As for $\overrightarrow{\beta}_2$,
there is no nice analytical expression,  but we can nevertheless  constrain it by conservation of the total missing 4-momentum $\slashchar{p}^\mu$,
\begin{equation}\label{pteqn}
   \slashchar{p}^\mu = \mathbf{\Lambda}_1^{-1}\left(
  \begin{array}{c}
    m_{\widetilde{\chi}_{1}^0}  \\
   \overrightarrow{0} \\
  \end{array}
\right)
+
 \mathbf{\Lambda}_2^{-1}\left(
  \begin{array}{c}
   m_{\widetilde{\chi}_{1}^\pm} - E''  \\
    - \overrightarrow{p}'' \\
  \end{array}
\right) ~~~~~~~~ \left[ \left(
  \begin{array}{c}
     E''  \\
     \overrightarrow{p}'' \\
  \end{array}
\right) \equiv \mathbf{\Lambda}_2 \left(
  \begin{array}{c}
     E_{\mu^\pm}  \\
     \overrightarrow{p}_{\mu^\pm} \\
  \end{array}
\right) \right]
\end{equation}
(ie. total missing 4-momentum = 4-momentum of one LSP plus 4-momentum of other LSP+$\nu$ system) of which the two transverse components $\overrightarrow{\slashchar{p}_T}$ are measurable,
in addition to the three kinematic constraints
\begin{equation}\label{muenergy}
    \overrightarrow{p}' \cdot \overrightarrow{p}'' = 0
     ~~~~,~~~~~ E'' = \frac{m_{\widetilde{\chi}_{1}^\pm}^2 - m_{\widetilde{\chi}_{1}^0}^2  }{2 m_{\widetilde{\chi}_{1}^\pm}}
     ~~~~,~~~~M_{e^+ e^-}^{max} = m_{\widetilde{\chi}_{2}^0} - m_{\widetilde{\chi}_{1}^0}
\end{equation}
Thus, (\ref{pteqn}) and (\ref{muenergy}) compose a system of five equations for the six unknowns
$\{ \overrightarrow{\beta}_2,$ $m_{\widetilde{\chi}_{1}^0},~ m_{\widetilde{\chi}_{2}^0},~m_{\widetilde{\chi}_{1}^\pm}\}$.
 If, as in a wide class of SUSY spectra (here controlled by SUSY parameters $\mu$, $M_{1,2}$, and $tan\beta$) we have the approximate relation $m_{\widetilde{\chi}_{1}^\pm} \approx m_{\widetilde{\chi}_{2}^0}$, then we are less one unknown and the five equations can be numerically solved for the masses $m_{\widetilde{\chi}_{1}^0}$ and $ m_{\widetilde{\chi}_{2}^0}$.

\section{Monte Carlo Test}
\label{sec:mc}

In a real experiment, there are of course many reasons why even a `perfect' event like Fig.~\ref{fig:3bodybeta} does not suffice for reliable mass reconstruction: measurement errors as well as inherent finiteness of detector resolution and sparticle widths will throw off the solution. Then there is the reality that no event is perfectly situated at an endpoint and, moreover, competition from backgrounds is expected. What we must do in practice, therefore, is to collect a number of events in some optimized neighborhood of the region of interest on the ``$M_{\ell^+ \ell^-}$ vs.  $\overline{M}_{{l2l}}$" plot, impose conditions (\ref{betaeqn})-(\ref{muenergy}) on each event (with $M_{e^+ e^-}^{max} \rightarrow M_{\ell^+ \ell^-}$ plus the assumption $m_{\widetilde{\chi}_{1}^\pm} \approx m_{\widetilde{\chi}_{2}^0}$ ), and study the distribution of extracted masses $m_{\widetilde{\chi}_{1,2}^0}$
(the reader is referred to the Appendix for a more detailed discussion of the numerical solving procedure).

Let us see how this might work by running the above programme through a Monte Carlo simulation of LHC data. Suppose for definiteness that Nature has chosen the Split SUSY parameter point considered in\cite{Kilian} with GUT-scale parameters
\begin{eqnarray*}
&  M_1 = M_2 = M_3 = 120\, \hbox{GeV}  &\\ \nonumber
& \mu = -90\, \hbox{GeV}      &\\ \nonumber
& tan \beta = 4  &
\end{eqnarray*}
in addition to a symmetry-breaking scale $\tilde{m} = 10^9\, \hbox{GeV}$. Integrating down to EW energies ($Q = m_Z$), all SUSY particles decouple except for the gluino and EW-inos, which attain the spectrum shown in Table~1; this is
 consistent with LEP and dark-matter constraints. At LHC energies the dominant chargino-neutralino production channels would then be\footnote{See\cite{Kilian} for a complete table of these.} $\widetilde{\chi}_{1}^\pm \widetilde{\chi}_{2}^0$ ($\sigma = 4650~fb$) and $\widetilde{\chi}_{1}^\pm \widetilde{\chi}_{3}^0$ ($\sigma = 2099~fb$), while the main neutralino-neutralino channel is $ \widetilde{\chi}_{2}^0 \widetilde{\chi}_{3}^0$ ($\sigma = 876~fb$).

\begin{table}
 \caption{\small \emph{Relevant masses (in GeV) at the Split-SUSY point we consider.}}
    \begin{center}
     \begin{tabular}{|c|c|c|c|c|c|c|} \hline
   ${\widetilde\chi}^0_1$
 &  ${\widetilde\chi}^0_2$
 &  ${\widetilde\chi}^0_3$
 &  ${\widetilde\chi}^0_4$
 &  ${\widetilde\chi}^\pm_1$
 &  ${\widetilde\chi}^\pm_2$
 &  $\tilde{g}$
 \\
 \hline
 $71.1$  & $109.9$ & $141.7$ & $213.7$ & $114.7$ & $215.7$ & $807.0$ \\ \hline
       \end{tabular}
    \end{center}
 \label{tab:mass}
\end{table}

LHC EW-ino events
($pp \to \widetilde{\chi}_{i}^0 \widetilde{\chi}_{j}^0, ~\widetilde{\chi}_{1,2}^\pm \widetilde{\chi}_{k}^0,~
 \widetilde{\chi}_{1,2}^\pm \widetilde{\chi}_{1,2}^\mp$)
  and SM backgrounds $ZZ$, $WZ$ and $W\gamma^*$ (see\cite{Vandelli} and\cite{Sullivan} for a good discussion of these and others not necessary for this study),
 corresponding to $300~fb^{-1}$ integrated luminosity are then generated via the HERWIG~6.5 package\cite{HERWIG65} and run through a simplified
detector simulator\footnote{The set-up is the same as in several of the author's previous publications (e.g.\cite{Huang1,Bian,Bisset2,Bisset4l}),
and includes a privately-coded  fast detector response simulation incorporating all the requisite simplified-geometry calorimetry, missing energy reconstruction, lepton isolation, etc., and has been checked against results in the literature using publicly available codes.
}.
 The following cuts are then employed, depending on the number of final leptons:

\vskip 0.2cm
\noindent
\underline{\bf For 2-Lepton Endstates}:
 \begin{itemize}
   \item Leptons must be isolated: no tracks of other charged particles are present in a $r = 0.3\, \hbox{rad}$ cone around the lepton, with less than $3\, \hbox{GeV}$ of energy deposited into the
electromagnetic calorimeter for $0.05\, \hbox{rad} < r < 0.3\, \hbox{rad}$
around the lepton.
   \item Leptons must be sufficiently hard:  $p_T^\ell > 10, 8\, \hbox{GeV}$ for $\ell = e,\mu$.
 \end{itemize}

 \vskip 0.2cm
\noindent
\underline{\bf For 3- and 4- Lepton Endstates}:
  \begin{itemize}
  \item Leptons must be hard and isolated as for 2-lepton endstates.
   \item Sufficient missing energy must be present in each event: $\slashchar{E_{T}} > 50\, \hbox{GeV}$.
 \end{itemize}
\textit{(note: 4-lepton selection criteria are shown for completeness only; the present study does not consider them further due to smallness of rate at \textbf{this} Split SUSY point)}

\vskip 0.2cm

\begin{figure}[!htb]
\begin{center}
\includegraphics[width=2.5 in]{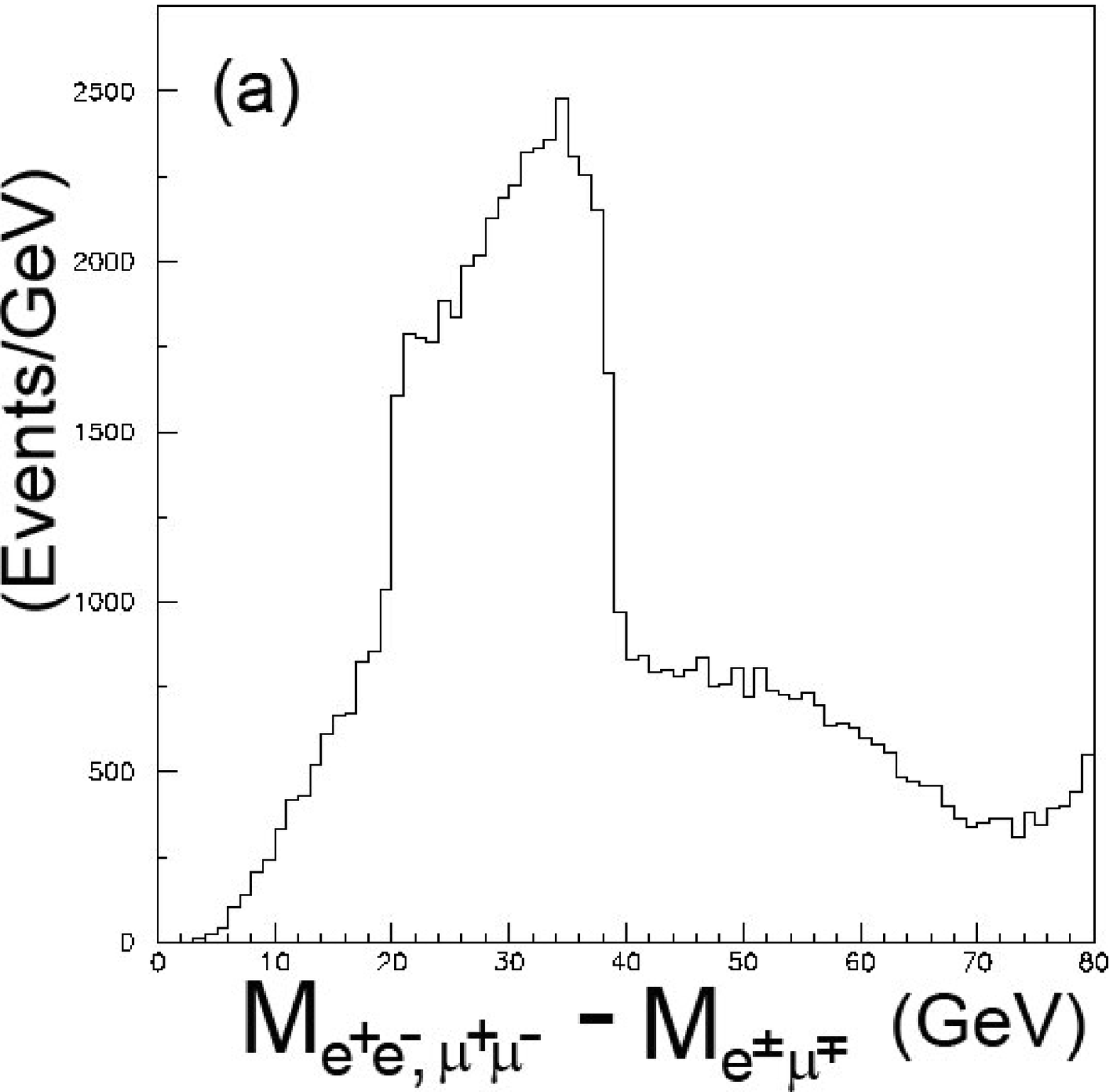}
\includegraphics[width=2.5 in]{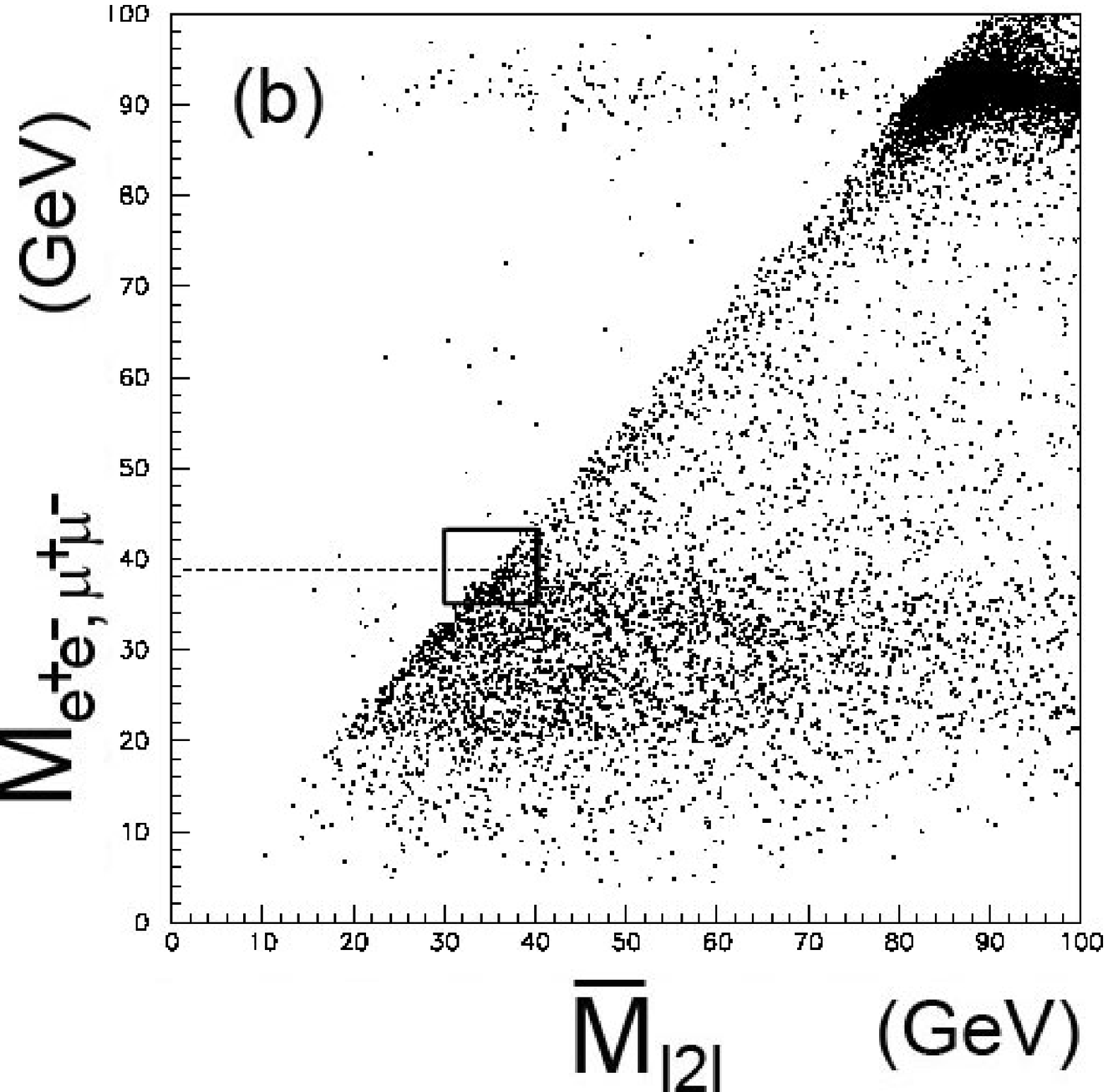}
\end{center}
\caption{\small \emph{ Invariant mass plots for $300~fb^{-1}$ luminosity (SUSY + SM):
(a) The flavor-subtracted dilepton invariant mass distribution clearly identifies the endpoint at
$\sim 39 \,\hbox{GeV}$.
(b) correlation between $M_{\ell^+ \ell^-}$ and $\overline{M}_{l2l}$;
events are taken in the boxed region shown, i.e. the neighborhood of where the line $M_{\ell^+ \ell^-} = 39\,\hbox{GeV}$
hits the envelope.}
}
 \label{fig:300fb}
\end{figure}

Though SM backgrounds are substantially reduced by these cuts, they still tend to far outnumber the SUSY signal events for 2- and 3-lepton endstates. We will see shortly there is no cause for worry, however, since SM backgrounds populate uninteresting regions of the relevant
invariant mass plots, shown in Fig.~\ref{fig:300fb}.

First observe the large number of 2-lepton events in Fig.~\ref{fig:300fb}a: there are roughly $6 \cdot 10^4$ SUSY $\ell^+ \ell^-$ events plus $3 \cdot 10^5$ SM events (mostly from Z decays and hence sitting near the Z-pole, $M_{\ell^+ \ell^-} \sim 91 \pm 10\,\hbox{GeV}$, not shown in the Figure) which, after subtracting wrong-flavor $e^\pm \mu^\pm$ combinations ($5 \cdot 10^4$ of these total), give us a  dilepton invariant mass distribution that clearly identifies an endpoint at $M_{\ell^+ \ell^-} \sim 39\,\hbox{GeV}$ due to $\widetilde{\chi}_{2}^0 \to \widetilde{\chi}_{1}^0$ decays\footnote{In this feasibility-level study, it is sufficient to mark this endpoint to within a few GeV; more complete analyses\cite{atlas,cms} of dilepton distributions with comparable or lower statistics verify harmlessness of SM backgrounds and suggest sub-GeV level precision is easily attainable.} (a second endpoint from $\widetilde{\chi}_{3}^0 \to \widetilde{\chi}_{1}^0$ decays near $\sim 70\,\hbox{GeV}$ is barely discernible but might be claimed statistically significant via a more in-depth analysis, e.g.\cite{Mohr}).\footnote{There is also an  `edge' at $M_{\ell^+ \ell^-} \sim 20\,\hbox{GeV}$, but this is actually
an effect of lepton pT cuts: since two leptons with momenta $p_\pm$ and relative angle $\theta$ give an invariant mass of $M_{\ell^+ \ell^-} = \sqrt{2 p_+ p_- (1 - \cos \theta)}$, cutting away $p_\pm < 10\,\hbox{GeV}$ will tend to deplete the $M_{\ell^+ \ell^-}$ spectrum below $2 p_\pm = 20\,\hbox{GeV}$.
}

\begin{figure}[!htb]
\begin{center}
\includegraphics[width=2.5 in]{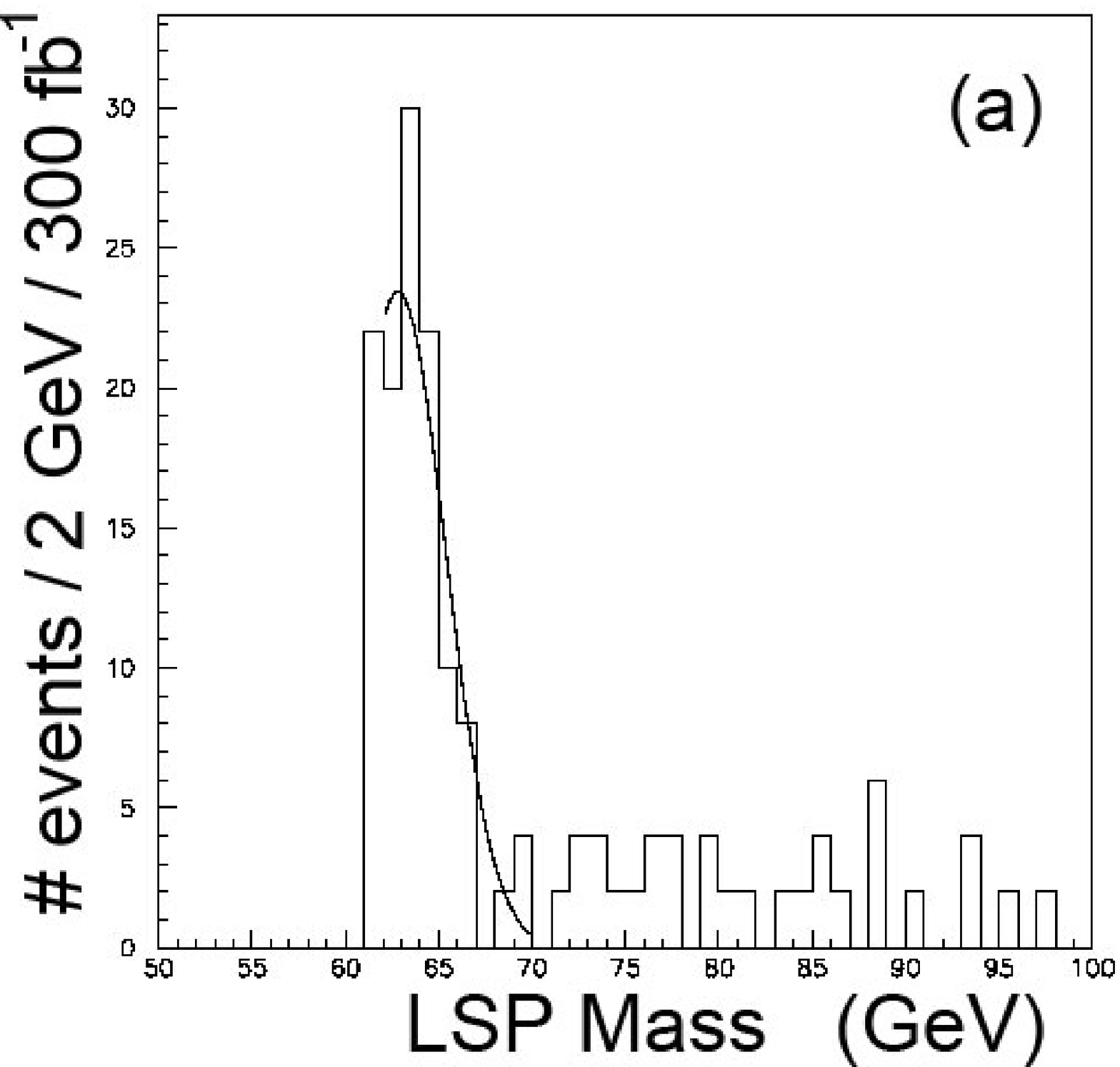}
\includegraphics[width=2.5 in]{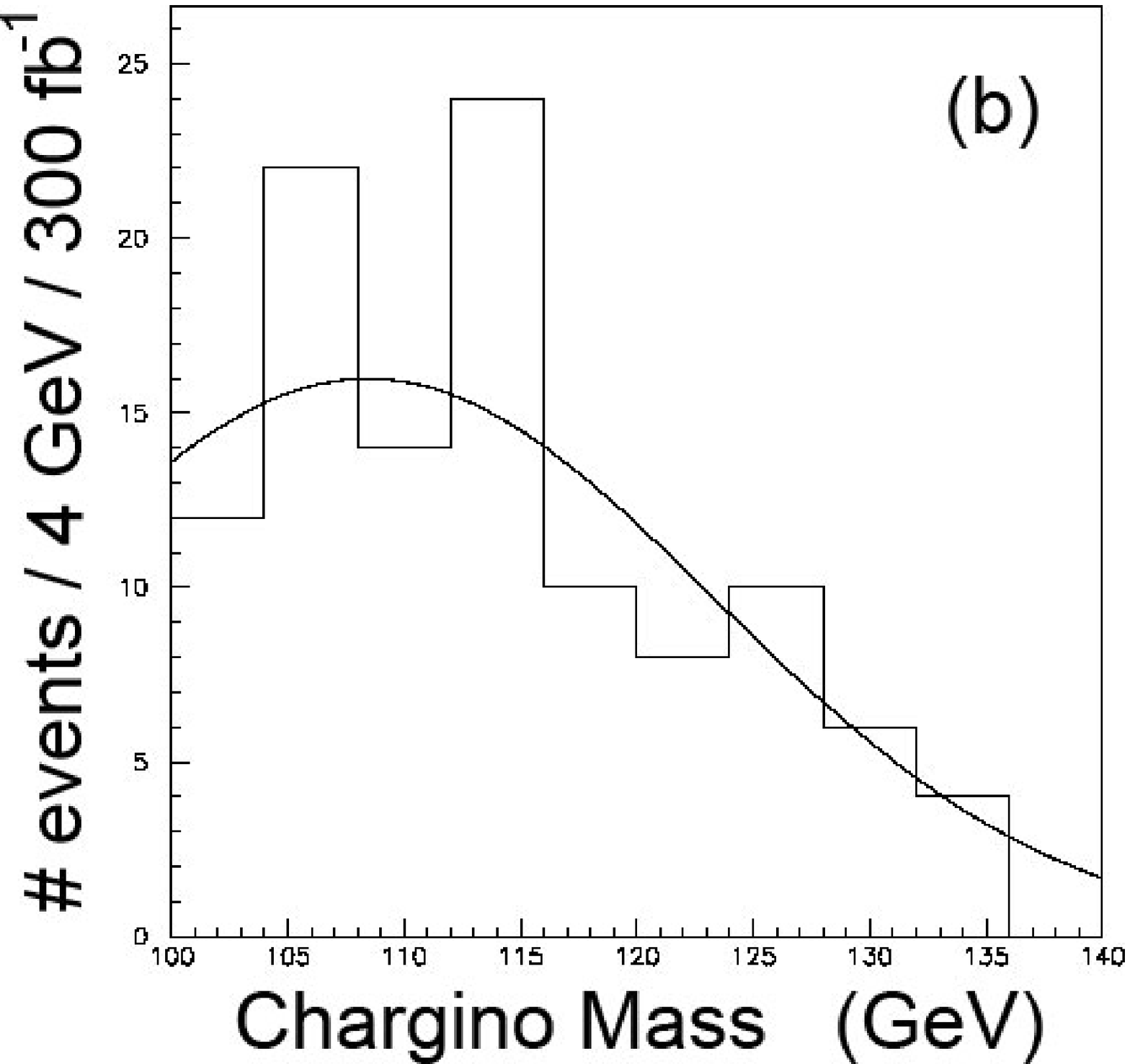}
\end{center}
\caption{\small \emph{ (a) Distribution of $m_{\widetilde{\chi}_{1}^0}$ from
events in the boxed region of Fig.~\ref{fig:300fb}b, assuming $m_{\widetilde{\chi}_{1}^\pm} = m_{\widetilde{\chi}_{2}^0}$. (b) Distribution of $m_{\widetilde{\chi}_{1}^\pm}$ assuming the correct value of $m_{\widetilde{\chi}_{1}^0} = 71\,\hbox{GeV}$.
In both plots the LEP2 constraint $m_{\widetilde{\chi}_{1}^\pm} \ge 100\,\hbox{GeV}$\cite{Benelli} has been applied.}
}
 \label{fig:m1}
\end{figure}

Turning now to 3-lepton events, SUSY  events ($e^+ e^- \mu^\pm + \mu^+ \mu^- e^\pm$ as well as same-flavor $e^+ e^- e + \mu^+ \mu^- \mu^\pm$ events\footnote{For same-flavor events, both possible lepton-pairings are plotted.}) number close to $\sim 8000$ against a SM background several times larger, but this latter is, in the `$M_{\ell^+ \ell^-}~vs.~\overline{M}_{l2l}$' plot shown in Fig.~\ref{fig:300fb}b,
 concentrated mostly up near the Z-pole again, and to a much lesser extent  throughout the bulk of the plot. The kinematically allowed region has a fairly clear diagonal `left edge', and it is where  the line $M_{\ell^+ \ell^-} = 39\,\hbox{GeV}$
 hits this edge that we should expect events of the type in Fig.~\ref{fig:3bodybeta}.
 Events are therefore collected from the boxed region shown, the limits of which ($M_{\ell^+ \ell^-} = 39 \pm 4\,\hbox{GeV}$, $\overline{M}_{l2l} < 40\,\hbox{GeV}$) give the optimal distribution of $m_{\widetilde{\chi}_{1}^0}$, shown in Fig.~\ref{fig:m1}a. Although this region certainly includes a large number of background events (e.g. $\widetilde{\chi}_{2}^0 \widetilde{\chi}_{3}^0 \to e^+e^- \mu^+ \mu^- 2 \widetilde{\chi}_{1}^0$, with one of the four leptons failing the hardness cut, or $\widetilde{\chi}_{2}^\pm \widetilde{\chi}_{2}^\mp(\to e^+e^- \mu^\mp \nu \widetilde{\chi}_{1}^0)$, SM processes $W \gamma^*$, as well as same-flavor signal events with the wrong lepton-pairing), their presence can be tolerated since these either do not give a physically-acceptable solution to the system of equations (\ref{betaeqn})-(\ref{muenergy}), hence are  rejected, or they give no \emph{preferred} solution and lead to a uniform `noise' in the  $m_{\widetilde{\chi}_{1}^0}$-distribution.
This latter, in fact, peaks sharply (a Gaussian fit gives $m_{\widetilde{\chi}_{1}^0} \sim 63 \pm 3\,\hbox{GeV}$) but somewhat lower (by about $15\% $) than the nominal value
in Table~1 (the $m_{\widetilde{\chi}_{2}^0}$-distribution looks the same but shifted by $\sim 39\,\hbox{GeV}$).
This deviation may arise from the fact that the assumption
$m_{\widetilde{\chi}_{1}^\pm} = m_{\widetilde{\chi}_{2}^0}$ is inaccurate by several percent at this parameter point (see Appendix for a discussion).
If, instead, one does not make this assumption, but knows beforehand the value of $m_{\widetilde{\chi}_{1}^0}$ from other measurements, (\ref{betaeqn})-(\ref{muenergy}) can also be used to find $m_{\widetilde{\chi}_{1}^\pm}$. The distribution  shown in
Fig.~\ref{fig:m1}b gives the correct $m_{\widetilde{\chi}_{1}^\pm}$ within errors ($m_{\widetilde{\chi}_{1}^\pm} = 108 \pm 15\,\hbox{GeV}$). Note this value of $m_{\widetilde{\chi}_{1}^\pm}$ can be put back into (\ref{betaeqn})-(\ref{muenergy}) to solve for  $m_{\widetilde{\chi}_{1}^0}$ again, iterating the process.

\section{Discussion and Conclusion}
\label{sec:disc}
This paper has introduced a Kinematic Selection technique applicable to EW-ino pair decays where 3- or 4-lepton events with specific parent-frame kinematics are captured for analysis.
Such a technique is particularly suited to Split SUSY models, where EW-inos are the only low-lying states expected to be observable at the LHC.
At the specific  parameter point studied,  the rate of $\widetilde{\chi}_{1}^\pm \widetilde{\chi}_{2}^0$-pair decays to
three leptons was sufficient to capture $O(10^2)$ interesting events, the kinematics of which could be reconstructed well enough to extract two EW-ino masses $m_{\widetilde{\chi}_{1,2}^0}$ within $15\%$ or so of their nominal values, assuming
 that $m_{\widetilde{\chi}_{1}^\pm} = m_{\widetilde{\chi}_{2}^0}$.
 It is, of course, entirely possible that other Split-SUSY parameter points would give higher rates
and thus allow us to study ${\widetilde{\chi}_{2}^0} {\widetilde{\chi}_{3,4}^0}$ modes as well as those involving the heavier chargino  $\widetilde{\chi}_{2}^\pm$ (here the kinematics are more complicated, but the final state also contains more leptons, hence more useful invariant mass constraints). Also, if EW-inos decay through an on-shell Z or W, or through a light Higgs boson, we can  in principle  look for leptonic invariant mass correlations which isolate events with specific parent-frame kinematics (note final state jets can also be included in this formalism); the method is quite flexible.

What other mass reconstruction methods are available for analyzing Split-SUSY EW-ino decays?
First consider the neutralinos.
There are now an array of methods which take advantage of the pair-production of neutralinos.
One class of ``Mass-Shell Techniques" (MST), represented in the work of\cite{Kawagoe} and\cite{Cheng},
essentially depends on maximizing the solvability of assumed mass-shell constraints in a given sample of events. This
seems quite effective for on-shell decays\footnote{But see\cite{Bisset} for some important caveats.}, but for the off-shell decay topologies  in the present work these methods
cannot be applied since there are not enough such constraints.
Recently fashionable ``transverse mass variable" methods\cite{mT2,mtvar}, e.g. $m_{T2}$, might be applied, though
these are usually stated for symmetric decays. In one such development\cite{Barr}, for example,
 a ``constrained mass variable" $m_{2C}$ proves quite powerful for
$\widetilde{\chi}_{2}^0 \widetilde{\chi}_{2}^0$ modes (followed by off-shell decays such as (\ref{z2decay}));
though such modes are expected to be negligible in Split-SUSY scenarios,
presumably $m_{2C}$ could be applied to the case of \emph{unlike} neutralinos $\widetilde{\chi}_{2}^0 \widetilde{\chi}_{3}^0$ as well. It's worth mentioning here that in the some of the latest developments with $m_{T2}$, e.g. $m_{T2}$-Assisted-On-Shell (MAOS) reconstruction\cite{maos}, information on the full LSP 4-momentum can be gleaned for both mass and spin determination.

 As for decay modes with charginos such as  $\widetilde{\chi}_{1}^\pm \widetilde{\chi}_{2}^0$, the author is not aware of any work showing how to reconstruct all the unknown masses (MAOS has not yet been tested\cite{maos-comm}) --- perhaps the above techniques can encompass these modes as well, but
there may be fundamental difficulties with extra invisible particles (neutrinos) in the decay products (e.g. MSTs would
have too many unknown degrees of freedom in each event).
  Finally, there is the  under-addressed question of multiple competing decay channels, e.g. when several different $\widetilde{\chi}_{i}^0 \widetilde{\chi}_{j}^0$ and $\widetilde{\chi}_{1}^\pm \widetilde{\chi}_{j}^0$ occur with
similar rates. The  case of $\widetilde{\chi}_{i}^0 \widetilde{\chi}_{j}^0$ yielding a $e^+e^- \mu^+ \mu^-$ endstate, in particular, is subject to a wedgebox analysis\cite{Bisset2} to partially separate events according to decay topology, though this has not been (but should be) extensively tested  for mass reconstruction methods which have so far only concentrated on a single channel. Note that in the method of the current paper this separation is unnecessary, since minima such as  $\overline{M}_{l2l}^{min}$  for various channels lie on different points of the envelope in Fig.~\ref{fig:300fb}b.
It seems quite natural that a combination of several techniques will be necessary to both isolate relatively pure samples of a given decay and reconstruct unknown masses as best as can be done at the LHC. For example, an MST-analysis might be applied to 4-lepton events from $\widetilde{\chi}_{2}^0 \widetilde{\chi}_{3}^0$ modes to get a ballpark estimate of
$ m_{\widetilde{\chi}_{1}^0}$, this value then used in the current method with $\widetilde{\chi}_{1}^\pm \widetilde{\chi}_{2}^0$ modes to determine $m_{\widetilde{\chi}_{1}^\pm}$, as done in Fig.~\ref{fig:m1}b.

In conclusion, then, this work represents the first application of a Kinematic Selection technique,
 found to be of particular use in Split-SUSY models. The strengths of Kinematic Selection include simplicity (relativistic kinematics) and robustness (works for multiple decay channels, even with backgrounds), which should make it a useful tool to experimentalists unraveling data from the LHC.

\section*{Acknowledgments}
This work was funded in part by the Kavli Institute for Theoretical Physics (Beijing). Thanks to A. Cohen for
useful discussion relevant to the manuscript.

\section*{Appendix}

\subsection*{Derivation of Minima}
In Section \ref{sec:app} the threshold minima (\ref{ml2lmin}) and  (\ref{4lmins}) were stated without derivation; here let us see how these were obtained.

Starting with $\overline{M}_{l2l}$ for a $e^+ e^- \mu^\pm$ endstate, choose the threshold frame of reference to be such that the electron/positron (of maximal invariant mass) are emitted along the z-axis $\pm \hat{z}$, while the muon (with maximal energy) is produced at spherical angles $\theta,~\phi$:
\begin{equation*}
p^\mu_{e^\pm} =  \left(
                       \begin{array}{c}
                         E \\
                        0  \\
                        0 \\
                         \pm E \\
                       \end{array}
                     \right)~, ~~~~  p^\mu_{\mu^\pm} =
                     \left(
                       \begin{array}{c}
                         E'' \\
                        E'' \sin\theta \cos\phi \\
                        E'' \sin\theta \sin\phi \\
                          E'' \cos\theta\\
                       \end{array}
                     \right)~ ~~
                       \left[ E \equiv \frac{m_{\widetilde{\chi}_{2}^0} - m_{\widetilde{\chi}_{1}^0}}{2} ~~,~~~
                        E'' \equiv \frac{m_{\widetilde{\chi}_{1}^\pm}^2 - m_{\widetilde{\chi}_{1}^0}^2  }{2 m_{\widetilde{\chi}_{1}^\pm}} \right]
\end{equation*}
Plugging these four-vectors into the definition of $\overline{M}_{l2l}$ in (\ref{avl2l}) and simplifying a bit, one obtains
\begin{eqnarray*}
 \overline{M}_{l2l} = \sqrt{\frac{m_{\widetilde{\chi}_{2}^0} - m_{\widetilde{\chi}_{1}^0}}{\sqrt{3} m_{\widetilde{\chi}_{1}^\pm}}}( 2 m_{\widetilde{\chi}_{1}^0}^4 - 2 m_{\widetilde{\chi}_{1}^0}^3 m_{\widetilde{\chi}_{1}^\pm} +
     m_{\widetilde{\chi}_{1}^0}^2 m_{\widetilde{\chi}_{1}^\pm}(2 m_{\widetilde{\chi}_{2}^0} - m_{\widetilde{\chi}_{1}^\pm}) + 2 m_{\widetilde{\chi}_{1}^0} m_{\widetilde{\chi}_{1}^\pm}^2 ( m_{\widetilde{\chi}_{1}^\pm} - 3 m_{\widetilde{\chi}_{2}^0}) &\\
 + m_{\widetilde{\chi}_{1}^\pm}^2 ( 2 m_{\widetilde{\chi}_{1}^\pm} + 3 m_{\widetilde{\chi}_{2}^0} - 2 m_{\widetilde{\chi}_{2}^0}m_{\widetilde{\chi}_{1}^\pm} ) + ( m_{\widetilde{\chi}_{1}^\pm} - m_{\widetilde{\chi}_{1}^0})^2 \cos 2\theta)^{1/4} \\
\end{eqnarray*}
This is clearly minimal when $\theta = \pi/2$, and further algebraic simplification leads to (\ref{ml2lmin}).

With a four-lepton endstate like $e^+ e^- \mu^+ \mu^-$, where the $e^+ e^- $ pair, say, has maximal invariant mass along the z-axis, one muon ($\mu^+$) will be going at an angle $\theta_+$ to the z-axis, while the other ($\mu^-$) has its own spherical angles $\theta_-,~\phi$:
\begin{equation*}
p^\mu_{e^\pm} =  \left(
                       \begin{array}{c}
                         E \\
                        0  \\
                        0 \\
                         \pm E \\
                       \end{array}
                     \right)~, ~~~~
                      p^\mu_{\mu^+} =
                     \left(
                       \begin{array}{c}
                         E_+ \\
                        E_+ \sin\theta_+ \\
                        0 \\
                          E_+ \cos\theta_+\\
                       \end{array}
                     \right)
                     , ~~~~
                      p^\mu_{\mu^-} =
                     \left(
                       \begin{array}{c}
                         E_- \\
                        E_- \sin\theta_- \cos\phi \\
                        E_- \sin\theta_- \sin\phi \\
                          E_- \cos\theta_-\\
                       \end{array}
                     \right)
\end{equation*}
where the energies $E_\pm$ take on a range of values determined by the relative angle between the muons and 3-body kinematics.
One \emph{could} then plug these expressions into the definitions of, e.g.,  $M_{4l}$ and $\overline{M}_{2l2l}$ and minimize over the angles, but it's much faster to intuit that since we're interested in minimizing an invariant mass, the muons should be going in the same direction ($\theta_+ = \theta_-$ and $\phi = 0$) which forces
$E_+ = E_- = (m_{\widetilde{\chi}_{3}^0}^2 - m_{\widetilde{\chi}_{1}^0}^2)/2m_{\widetilde{\chi}_{3}^0}$; plugging this into the definitions (\ref{4linv}) and simplifying yields the quoted minima (\ref{4lmins}), and numerically sampling over $(\theta_+,\theta_-,\phi)$-space confirms these are indeed correct.

\subsection*{Numerically Solving for the Masses}

The most direct method of numerically solving the system of  equations (\ref{pteqn}) and (\ref{muenergy}) for the variables $\{ \beta_{2x},~\beta_{2y},~\beta_{2z},~$ $m_{\widetilde{\chi}_{1}^0},~ m_{\widetilde{\chi}_{2}^0},~m_{\widetilde{\chi}_{1}^\pm}\}$ (where, say, $m_{\widetilde{\chi}_{1}^0}$ is known ) is to simply loop over a liberal range of values for $m_{\widetilde{\chi}_{1}^\pm}$ (say between $50\,\hbox{GeV}$ and $300\,\hbox{GeV}$ ) and the components of $\overrightarrow{\beta}_2$ ($\beta_{2{x,y,z}}$ each between -1 and 1, with $|\overrightarrow{\beta}_2| < 1$), imposing the other constraints inside these four nested loops.
In fact, we only need to loop over two of the components of $\overrightarrow{\beta}_2$, since the third is fixed by a requirement on the transformed muon energy: when the muon four-momentum $p_\nu$ is boosted back to the chargino's decay frame by $\overrightarrow{\beta}_2$,  i.e.
\begin{equation} \nonumber
\left(
  \begin{array}{c}
    E'' \\
    p''_{ x} \\
    p''_{ y} \\
   p''_{ z} \\
  \end{array}
\right) =
\left(
  \begin{array}{cccc}
    \gamma_2 & \beta_{2x} \gamma_2 & \beta_{2y} \gamma_2 & \beta_{2z} \gamma_2 \\
   \beta_{2x} \gamma_2 & 1 + (\gamma_2-1)\frac{\beta_{2x}^2}{\beta_2^2} & (\gamma_2-1)\frac{\beta_{2x} \beta_{2y}}{\beta_2^2} & (\gamma_2-1)\frac{\beta_{2y}\beta_{2z}}{\beta_2^2} \\
    \beta_{2y} \gamma_2 & (\gamma_2-1)\frac{\beta_{2x} \beta_{2y}}{\beta_2^2} & 1+ (\gamma_2-1)\frac{\beta_{2y}^2}{\beta_2^2} & (\gamma_2-1)\frac{\beta_{2y} \beta_{2z}}{\beta_2^2} \\
    \beta_{2z} \gamma_2 &(\gamma_2-1)\frac{\beta_{2x} \beta_{2z}}{\beta_2^2} &  (\gamma_2-1)\frac{\beta_{2y} \beta_{2z}}{\beta_2^2} & 1+ (\gamma_2-1)\frac{\beta_{2z}^2}{\beta_2^2} \\
  \end{array}
\right)
\left(
  \begin{array}{c}
    E \\
    p_{ x} \\
    p_{ y} \\
   p_{ z} \\
  \end{array}
\right)
\end{equation}
we must satisfy the constraint from 3-body kinematics,
\begin{equation}\label{muenergy2} \nonumber
    E'' = \frac{m_{\widetilde{\chi}_{1}^\pm}^2 - m_{\widetilde{\chi}_{1}^0}^2  }{2 m_{\widetilde{\chi}_{1}^\pm}}
    = \frac{E - \beta_{2x} p_x - \beta_{2y} p_y - \beta_{2y} p_y}{\sqrt{1 -\beta_{2x}^2 -\beta_{2y}^2 -\beta_{2z}^2}}
\end{equation}
This can be rearranged into a quadratic equation for $\beta_{2x}$,
\begin{eqnarray}
0  &=&   A \beta_{2x}^2 + B \beta_{2x} + C   \\ \nonumber
\textrm{where} && \\\nonumber
    A &\equiv& 1 + (p_x / E'')^2 \\\nonumber
    B &\equiv& -2 (p_x / E'')(E/E'' - p_y/E'' \beta_{2y} - p_z/E'' \beta_{2z}) \\\nonumber
    C &\equiv&  \beta_{2y}^2 + \beta_{2z}^2 - 1 + (E/E'' - p_y/E'' \beta_{2y} - p_z/E'' \beta_{2z})^2 \\\nonumber
\end{eqnarray}
There is of course a potentially two-fold ambiguity in the solution for $\beta_{2x}$, and both values must be tried (if they are indeed real and satisfy $|\beta_2| < 1$).
Scanning over the 3-dimensional $(m_{\widetilde{\chi}_{1}^\pm}, \beta_{2y}, \beta_{2z})$-space then, we look for the point which best satisfies the missing-momentum constraints (these must be satisfied within $\pm 10 \,\hbox{GeV}$) as well as the kinematic constraint $\overrightarrow{p}' \cdot \overrightarrow{p}'' = 0$, achieved by minimizing $\Delta$:
\begin{eqnarray*}
  \Delta &\equiv& \sqrt{\Delta_1^2 + \Delta_2^2 + \Delta_3^2} \\
  \Delta_1 & \equiv & \slashchar{p}^x + \beta_{1x} \gamma_1 m_{\widetilde{\chi}_{1}^0} + \\
  &&
  \beta_{2x} \gamma_2 (m_{\widetilde{\chi}_{1}^\pm} - E'') - (1 + (\gamma_2-1)\frac{\beta_{2x}^2}{\beta_2^2})p''_x - (\gamma_2-1)\frac{\beta_{2x} \beta_{2y}}{\beta_2^2}p''_y - (\gamma_2-1)\frac{\beta_{2y}\beta_{2z}}{\beta_2^2}p''_z
   \\
  \Delta_2 & \equiv & \slashchar{p}^y + \beta_{1y} \gamma_1 m_{\widetilde{\chi}_{1}^0} + \\
  && \beta_{2y} \gamma_2 (m_{\widetilde{\chi}_{1}^\pm} - E'') - (1 + (\gamma_2-1)\frac{\beta_{2y}^2}{\beta_2^2})p''_y - (\gamma_2-1)\frac{\beta_{2x} \beta_{2y}}{\beta_2^2}p''_x - (\gamma_2-1)\frac{\beta_{2y}\beta_{2z}}{\beta_2^2}p''_z
  \\
  \Delta_3 & \equiv & \alpha ( \overrightarrow{p}' \cdot \overrightarrow{p}'')/(|\overrightarrow{p}'| |\overrightarrow{p}''|)
\end{eqnarray*}
where $\beta_1$ is already known from (\ref{betaeqn}), and $\alpha$ is a weight (high $\alpha \sim 1000$ seems best, meaning that all solutions have essentially perpendicular leptons, $|\cos \theta| < 0.01$): this minimization always gave a unique solution in all cases tested.
Moreover, this procedure is efficient in dealing with backgrounds (or same-flavor $e^+ e^- e^\pm$ and $\mu^+ \mu^- \mu^\pm$ events with the wrong lepton-pairing): either these fail to satisfy both missing energy constraints ($\Delta_{1,2}$) within $\pm 10\,\hbox{GeV}$, or yield solutions randomly distributed across mass space, which merely provides a uniform 'noise' in the solution histogram.

The same algorithm applies, of course, when $m_{\widetilde{\chi}_{1}^0}$ is not known (so we loop over \emph{it}) but we assume $m_{\widetilde{\chi}_{1}^\pm} = m_{\widetilde{\chi}_{2}^0} $. Here, however, since actual kinematic data comes from an event where this equality does not strictly hold, we expect the result of the numerical solution above to have a systematic error: setting $m_{\widetilde{\chi}_{1}^\pm}  = m_{\widetilde{\chi}_{2}^0} + \epsilon$, the numerical solution $m_{\widetilde{\chi}_{1}^0}'$ is offset from the actual value,  $m_{\widetilde{\chi}_{1}^0}' = m_{\widetilde{\chi}_{1}^0} + \delta$. To quantitatively understand the relationship between  $\delta$ and $\epsilon$, we would be best off running many simulations at different Split-SUSY parameter points and plotting the correlation `$\delta$-versus-$\epsilon$'; but since this is extremely time-intensive and not practical for the present work, we can get a quick-and-rough idea in the perturbative limit ($\delta,~\epsilon << m_{\widetilde{\chi}_{1}^0}$)  by letting $(m_{\widetilde{\chi}_{1}^0}',~ m_{\widetilde{\chi}_{1}^\pm}',~ m_{\widetilde{\chi}_{2}^0}' )$ and
$(m_{\widetilde{\chi}_{1}^0},~m_{\widetilde{\chi}_{1}^\pm},~m_{\widetilde{\chi}_{2}^0} )$ both solve $\Delta_1 = 0$ for the same Lorentz boosts, where
\begin{eqnarray} \label{approx}
 m_{\widetilde{\chi}_{1}^\pm} &=& m_{\widetilde{\chi}_{2}^0} + \epsilon \\ \nonumber
  m_{\widetilde{\chi}_{1}^0}' &=& m_{\widetilde{\chi}_{1}^0} + \delta \\ \nonumber
  m_{\widetilde{\chi}_{1}^\pm}' &=& m_{\widetilde{\chi}_{2}^0}' =   m_{\widetilde{\chi}_{2}^0} + \delta \\ \nonumber
\end{eqnarray}
and then seeing what the relationship between $\delta$ and  $\epsilon$  must be.
Thus, the exact and approximate solutions, respectively, satisfy
\begin{eqnarray} \label{sat1}
   0 & = & \slashchar{p}^x + \beta_{1x} \gamma_1 m_{\widetilde{\chi}_{1}^0} +
  \beta_{2x} \gamma_2 \frac{m_{\widetilde{\chi}_{1}^\pm}^2 - m_{\widetilde{\chi}_{1}^0}^2}{2 m_{\widetilde{\chi}_{1}^\pm}} + \Omega
   \\ \label{sat2}
 0 & = & \slashchar{p}^x + \beta_{1x} \gamma_1 m_{\widetilde{\chi}_{1}^0}' +
  \beta_{2x} \gamma_2 \frac{m_{\widetilde{\chi}_{1}^\pm}'^2 - m_{\widetilde{\chi}_{1}^0}'^2}{2 m_{\widetilde{\chi}_{1}^\pm}'} + \Omega \\ \nonumber
  \mathrm{where} \\ \nonumber
  \Omega & \equiv  & - (1 + (\gamma_2-1)\frac{\beta_{2x}^2}{\beta_2^2})p''_x - (\gamma_2-1)\frac{\beta_{2x} \beta_{2y}}{\beta_2^2}p''_y - (\gamma_2-1)\frac{\beta_{2y}\beta_{2z}}{\beta_2^2}p''_z
\end{eqnarray}
Explicitly inserting the $\delta$- and $\epsilon$-dependencies from (\ref{approx}) into (\ref{sat1}) and (\ref{sat2}) and setting these latter equal,
\begin{eqnarray*}
    \beta_{1x} \gamma_1 m_{\widetilde{\chi}_{1}^0} +
  \beta_{2x} \gamma_2 \frac{m_{\widetilde{\chi}_{1}^\pm}^2 - m_{\widetilde{\chi}_{1}^0}^2}{2 m_{\widetilde{\chi}_{1}^\pm}}
 & = &\beta_{1x} \gamma_1 m_{\widetilde{\chi}_{1}^0}' +
  \beta_{2x} \gamma_2 \frac{m_{\widetilde{\chi}_{1}^\pm}'^2 - m_{\widetilde{\chi}_{1}^0}'^2}{2 m_{\widetilde{\chi}_{1}'^\pm}} \\
  && \\
   \Rightarrow&& \\
   && \\
    \beta_{1x} \gamma_1 m_{\widetilde{\chi}_{1}^0} +
  \beta_{2x} \gamma_2 \frac{(m_{\widetilde{\chi}_{2}^0}+ \epsilon)^2 - m_{\widetilde{\chi}_{1}^0}^2}{2 (m_{\widetilde{\chi}_{2}^0}+ \epsilon)}
 & = &\beta_{1x} \gamma_1 (m_{\widetilde{\chi}_{1}^0}+\delta) +
  \beta_{2x} \gamma_2 \frac{(m_{\widetilde{\chi}_{2}^0}+ \delta)^2 - (m_{\widetilde{\chi}_{1}^0}+\delta)^2}{2 (m_{\widetilde{\chi}_{2}^0}+ \delta)}
\end{eqnarray*}
And now expanding and keeping only terms of $O(\delta)$ or $O(\epsilon)$, we finally arrive at
\begin{equation}\label{correlation}
    \delta \approx
    \left(\frac{1+r^2}
    {\frac{2 \beta_{1x} \gamma_1}{\beta_{2x} \gamma_2} +
    (1-r)^2         }\right)\epsilon ~~~~~~~~~~
     \left[ r \equiv \frac{m_{\widetilde{\chi}_{1}^0}}{m_{\widetilde{\chi}_{2}^0}}         \right]
\end{equation}
From (\ref{correlation}) we see that, even in this greatly simplified treatment, the sign and magnitude of the correlation between   $\delta$ and $\epsilon$ depends on the Lorentz boosts $\overrightarrow{\beta}_{1,2}$ specific to each event.
All things being equal, however, $\beta_{1x} \gamma_1 \approx -\beta_{2x} \gamma_2$ (the $\widetilde{\chi}_{2}^0$ and
$\widetilde{\chi}_{1}^\pm$ may tend to go in opposite directions), we can drop the smallish $r$-dependent terms, and we might thus expect a  negative correlation $\delta \approx -\epsilon \cdot$. This is indeed what is observed at the Split-SUSY point in the present study, where $\delta \approx -8 \,\hbox{GeV}$ and $\epsilon \approx 5 \,\hbox{GeV}$.

\end{document}